\documentclass[12pt,twoside,a4paper,english]{article}

\usepackage[english]{babel}
\usepackage{epsfig}
\usepackage{amscd}
\usepackage{amssymb}
\usepackage{mathrsfs}
\usepackage{amsfonts}
\usepackage{graphicx}
\usepackage{graphics}
\usepackage{amsmath}
\usepackage{bm}
\usepackage{overcite}
\usepackage{indentfirst}

\newlength{\dinwidth}
\newlength{\dinmargin}
\addtocounter{secnumdepth}{1}

\newcommand{\no}{\noindent}

\newcommand{\mean}[1]{\ensuremath{\left< #1 \right>}}

\begin{document}


\title{Probing Noise in Gene Expression\\ and Protein Production}

\date{}
\author
{Sandro Azaele,$^{1}$ Jayanth R. Banavar, $^{2}$ and Amos Maritan,$^{3}$ \\
\\
\normalsize{$^{1}$Department of Civil and Environmental Engineering,
E-Quad,}\\
\normalsize{Princeton University, Princeton, NJ 08544, USA.}\\
\normalsize{$^{2}$Department of Physics, The Pennsylvania State
University,}\\
\normalsize{104 Davey Laboratory, University Park, Pennsylvania 16802, USA.}\\
\normalsize{$^{3}$Dipartimento di Fisica ``Galileo Galilei,'' Universit\`{a} di Padova,}\\
\normalsize{CNISM-Unit\`{a} di  Padova and INFN, via Marzolo 8,
I-35131, Padova, Italy.}}

\maketitle

\textbf{We derive exact solutions of simplified models for the
temporal evolution of the protein concentration within a cell
population arbitrarily far from the stationary state. We show that
monitoring the dynamics  can assist in modeling and understanding
the nature of the noise and its role in gene expression and protein
production. We introduce a new measure, the cell turnover
distribution, which can be used to probe the phase of transcription
of DNA into messenger RNA.}\\


Advances in experimental techniques, that enable the direct
observation of gene expression in individual cells, have
demonstrated the importance of stochasticity in gene expression, the
translation into proteins of the information encoded within DNA
\cite{kaern,ras,paul1,paul2,xie1}. Such variability can lead to
deleterious effects in cell function and cause diseases \cite{mag}.
On the positive side, stochasticity in gene expression confers on
cells the ability to be responsive to unexpected stresses and may
augment growth rates of bacterial cells compared to homogeneous
populations \cite{that1}. Disentangling the various contributions to
production fluctuations is further complicated by the recent finding
that different stochastic processes yield the same response in the
variance in protein abundance at stationarity \cite{Pedraza}. A
population of isogenic cells growing under the same environmental
conditions can exhibit protein abundances that vary greatly from
cell to cell. The sources of variability have been identified at
multiple levels \cite{ros,ped,ras2,bec,elow}, with transcription and
translation playing a major role under certain circumstances
\cite{new,bar,mca}.

The low concentration of reactants potentially has two important
consequences: the first is that fluctuations around the mean can be
large; the second is that the nature of the stochastic noise should
be taken into account in some detail because one may not simply
invoke the central limit theorem \cite{Kam} which leads to the
universal and ubiquitous Gaussian noise. Thus, two genes expressed
at the same average abundance can produce populations with different
phenotypic noise strengths, defined as the ratio of the variance
over the mean value of the number of proteins \cite{ozb}. We show
here that two distinct models, one taking into account the detailed
nature of the noise and the other following from an application of
the central limit theorem, yield exactly the same stationary
solution for the distribution of proteins in isogenic cells under
the same environmental conditions. The exact dynamical solution of
these two simplified models demonstrate the value of monitoring the
dynamics for understanding the nature of the noise in a cell.

\begin{figure}[h]
\begin{center}
\includegraphics[width=10cm]{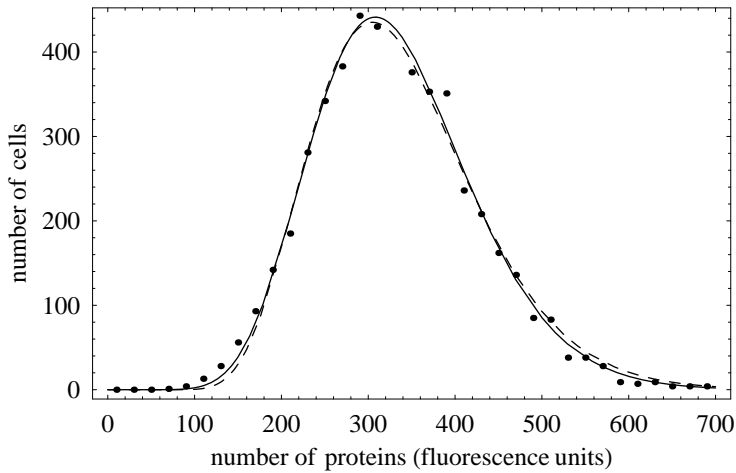}\\
\caption{The stationary distribution of proteins in a prokaryotic
cell population taken from Ref.\cite{ozb} fitted to Eq. (\ref{ssng})
with $\delta=0$
 or $\delta/\gamma_2=60.3$ (dashed). The best fit parameters are $\gamma_1/k_2=0.038$,
$k_1/\gamma_2=12.88$ ($\chi^2\simeq6100$) and $\gamma_1/k_2=0.030$,
$k_1/\gamma_2=8.33$ ($\chi^2\simeq8700$), respectively. From the
experimental data it is hard to distinguish between the steady state
distributions predicted by Eq. (\ref{ssng}) with $\delta=0$ and
$\delta>0$.}\label{prot}
\end{center}
\end{figure}

We make the simplified assumption that the kinetics of gene
expression can be described approximately by four rate constants:
$k_1$ and $k_2$ are the transcription and translation rates,
respectively and $\gamma_1$ and $\gamma_2$ are the degradation rates
for mRNA and proteins, respectively. It has been found
experimentally that proteins are produced in bursts
\cite{xie1,that2,ozb,kie} with an exponential distribution of the
number of proteins produced in a given event. Following Paulsson
\textit{et al.} \cite{paulsson} and Friedman \textit{et al.}
\cite{xie2}, we will assume that transcription pulses are Poisson
events and that the probability distribution that in a single event
$I>0$ proteins are produced, $w(I)$, is approximated by
$w(I)=\frac{\gamma_1}{k_2} e^{-\frac{\gamma_1}{k_2} I}$, where
$k_2/\gamma_1$ is the translation efficiency, i.e. the mean number
of proteins produced in a given burst. Here we consider a simple
model for the production of proteins without memory and aging of
molecules. Using the specific burst distribution given above allows
one to obtain the shape of the protein distribution even far from
stationarity. Under these assumptions, the stochastic equation that
governs the single-variable dynamics of gene expression, can be
written as

\begin{equation}\label{ngwn}
 \dot{x}(t)=\delta-\gamma_2\, x(t) +\Lambda(t).
\end{equation}

This pseudo-equation describes the real-time stochastic evolution of
gene expression through a deterministic part and a stochastic term
$\Lambda(t)$, which will be defined later on. Here $x$ is a
continuous variable that represents the number of proteins within a
cell. $\delta$ is a term added for generality which can be
incorporated in the average noise.

In order to understand the nature of the noise for the gene
expression case, let us consider the random variable, $I_k$, that is
a measure of the number of proteins in the $k^{\textrm{th}}$
transcription event, where $k=1,2,\ldots,n$. A key quantity of
interest is $\sum_{k=1}^{n(t)}I_k\equiv \Lambda(t) \Delta t$ where
$n(t)$, the number of events in the time interval $(t,t+\Delta t)$,
is a random variable independent of both $x$ and the $I_k$s. As in
the experiment, let us postulate that: \textit{i)} the $I_k$s are
independent and identically distributed with exponential
distribution; and \textit{ii)} the probability of $n$ events
occurring during the time interval $\Delta t$ is given by the
Poisson distribution $q_n(\Delta t)=(k_1\Delta t)^n\exp(-k_1\Delta
t)/n!$. The distribution of $\Lambda(t)$ that we use in Eq.
(\ref{ngwn}) can be explicitly calculated \cite{si} and leads to the
following expression for the cumulants: $\label{new20} \langle
\langle
\Lambda(t_1)\cdots\Lambda(t_n)\rangle\rangle=n!k_1\left(k_2/\gamma_1\right)^n
\prod_{i=2}^n\delta(t_i-t_{1})$ for $n\geq 2$ and
$\langle\Lambda(t)\rangle=k_1 k_2/\gamma_1$, independent of time.
Because the cumulants are delta functions, the noise is still white
(events are uncorrelated if they occur at different times); however
the noise is no longer Gaussian because cumulants with $n$ greater
than two are non-zero.

The master equation that describes this burst-like process is
\cite{Kam}
\begin{eqnarray}\label{intdiff}
\frac{\partial p(x,t)}{\partial t}&=&
 -\frac{\partial}{\partial x}[(\delta - \gamma_2\, x)p(x,t)]+\nonumber\\
 &+& k_1\int_0^{x}w(x-y)p(y,t)\textrm{d}y -k_1 p(x,t)\ ,
\end{eqnarray}
\no where $p(x,t)\equiv p(x,t|x_0,0)$ is the conditional probability
that the protein concentration has a value $x$ at time $t$ given
that it has a value $x_0$ at time $0$; and
$w(x)=\frac{\gamma_1}{k_2} e^{-\frac{\gamma_1}{k_2} x}$. The
stationary solution of this model (with $\delta=0$) was first
obtained by Paulsson et al. \cite{paulsson} and subsequently
re-derived by Friedman et al. \cite{xie2}. For arbitrary $\delta>0$,
we find the stationary solution is
\begin{equation}\label{ssng}
  p_{s}(x)=\left(\frac{\gamma_1}{k_2}\right)^{\frac{k_1}{\gamma_2}}
  \frac{\Theta(x-\delta/\gamma_2)}{\Gamma(k_1/\gamma_2)}
  \left(x-\frac{\delta}{\gamma_2}\right)^{\frac{k_1}{\gamma_2}-1}
  e^{-\frac{\gamma_1}{k_2}(x-\frac{\delta}{\gamma_2})}
\end{equation}
\no where $\Theta(x)$ is the step function equal to $1$ when $x>0$
and zero otherwise. This distinctive feature is a sharp signature of
the nature of the noise even in the stationary solution but is
present only when $\delta \neq 0$. However, as shown in the fit to
the stationary solution in Fig. (\ref{prot}), the singularity, if it
exists, is easily masked by other noise effects leading to a
rounding effect.

\begin{figure}[h]
\begin{center}
  \includegraphics[width=10cm]{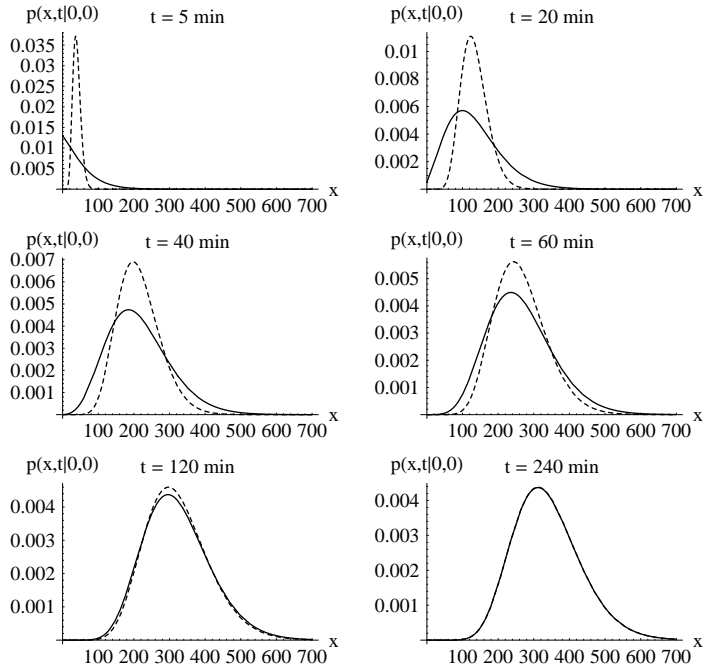}\\
  \caption{Protein distribution dynamics for different types
  of noise and with the same initial conditions, i.e. $x_0=0$ proteins
  at $t=0$. The dashed curve is for the multiplicative Gaussian noise,
  i.e. Eq. (\ref{gwn}) with $f\equiv k_1 k_2/\gamma_1$ and $D\equiv \gamma_2
k_2/\gamma_1$ \cite{si};
  whereas the other curve is for the non-Gaussian noise, i.e. for Eq. (\ref{gs2}).
  In both cases the parameters are $\delta=0$, $\gamma_2/D=\gamma_1/k_2=0.038$,
  $f/D=k_1/\gamma_2=12.88$ and we have set
  $\gamma_2^{-1}=40$ min, $\gamma_1^{-1}=2$ min.}\label{dyn}
\end{center}
\end{figure}

Although experiments on gene expression\cite{xie1,ozb} are
consistent with a burst-like protein production, steady-state
distributions of protein abundances are equally compatible with
alternative explanations. In fact, because mRNA is unstable compared
to protein lifetime ($\gamma_1\gg \gamma_2$), one can assume that
transcripts give rise to a constant flux of proteins $f$ and
subsequently any protein degrades at a constant rate $\gamma_2$.
Because of the great amount of available molecules, one can apply
the central limit theorem and suppose that the amplitude of
fluctuations is simply proportional to $\sqrt{x}$. Within this
framework there is no burst-like production, nevertheless the
stationary solutions that one obtains for a burst-like process,
including that of the extended autoregulation model \cite{bec2,isa},
are also obtained in models with appropriately chosen random
multiplicative Gaussian noise \cite{si}. Within this scenario the
stochastic evolution of the protein concentration $x(t)$ is governed
by the equation

\begin{equation}\label{gwn}
\dot{x}(t)=f-\gamma_2\, x(t) +\sqrt{Dx(t)}\eta(t),
\end{equation}

\no where $\eta(t)$ is a Gaussian white noise with autocorrelation
$\mean{\eta(t)\eta(t')}=2\delta(t-t')$. Note that the same equation
could be obtained on setting $\langle \Lambda(t)\rangle=f$ and
$\langle\langle\Lambda(t)\Lambda(t')\rangle\rangle\equiv
\langle\Lambda(t)\Lambda(t')\rangle -
\langle\Lambda(t)\rangle\langle\Lambda(t')\rangle=2Dx(t)\delta(t-t')
$ in Eq. (\ref{ngwn}) with all higher order cumulants being
identically zero. We point out that in ecology Eq. (\ref{gwn}) is
useful for studying the evolution of tropical forests \cite{aza},
where the detailed nature of the stochastic noise is not important
because of the relatively large numbers of trees of a given species.
In the field of finance, Eq. (\ref{gwn}) has been used to study the
evolution of interest rates (the Cox-Ingersoll-Ross model
\cite{cir}), where analogous considerations on fluctuations can be
made. On defining $f\equiv k_1 k_2/\gamma_1$ and $D\equiv \gamma_2
k_2/\gamma_1$, Eq. (\ref{gwn}) yields the same stationary state as
in Eq. (\ref{intdiff}) with $\delta=0$, i.e. Eq. (\ref{ssng}). The
mean number of proteins at stationarity is $k_1k_2/\gamma_1\gamma_2$
and the phenotypic noise strength at stationarity is $k_2/\gamma_1$,
relations that are consistent with previous findings \cite{ozb}. In
order to take into account the effects of feedback in a system
undergoing auto-regulation, one can introduce the physically
transparent modification $f \rightarrow Dc(x)$, where $c$ is a
response function which can be modeled as having two distinct
limiting values at zero and at infinity with the latter being
smaller than the former. Even in this situation, we obtain the same
stationary distribution with bistability as Friedman et al.
\cite{xie2}.
Despite this much more realistic analysis, the final stationary
protein distribution is experimentally indistinguishable from Eq.
(\ref{ssng}) with $\delta=0$. Thus, a theoretical modeling of the
stationary state of protein production provides little insight into
the microscopic nature of the noise that leads to stationarity.
According to our approach, stochasticity in gene expression ensues
from the large number of available components which entangle a lot
of different mechanisms within a cell. Interestingly, this calls for
effective mechanisms which can dampen the deleterious effects of
protein noise. Such efficient noise-reducing mechanisms could be a
combination of gestation and senescence, because of their ability to
prevent fluctuations rather than correcting \cite{Pedraza}.

These results raise the question whether the agreement between the
stationary solutions of the theoretical models and experiments are
in fact a direct probe of the nature of the microscopic noise and
whether the asymmetric stationary solutions derive from a careful
consideration of the bursty nature of the noise. In order to
circumvent the indistinguishability of steady-states, one can look
into empirical protein abundances far from stationarity, for which
we provide analytical formulas. Thus we turn now to a study of the
dynamics of Eq. (\ref{intdiff}) which is a powerful probe of the
noise effects. We have derived \cite{si} the solution at arbitrary
time,

\begin{eqnarray}\label{gs2}
p(x,t) & = & e^{-k_1 t}\ \delta(x-\xi_t)
  + \Theta(x-\xi_t)\ \frac{k_1\gamma_1}{k_2\gamma_2}
  \left(e^{\gamma_2 t}-1\right)e^{-k_1 t}\times\nonumber\\
   &\times& \exp\left[-\frac{\gamma_1}{k_2} e^{\gamma_2 t}
 (x-\xi_t)\right]
  {}_{1}F_1\left(\frac{k_1}{\gamma_2}+1,2;
  \frac{\gamma_1}{k_2} (e^{\gamma_2 t}-1)(x-\xi_t)\right),
\end{eqnarray}

\no where ${}_{1}F_1\left(a,b;x\right)$ is the confluent
hypergeometric function \cite{Leb} and $\xi(t)\equiv x_0e^{-\gamma_2
t}+\frac{\delta}{\gamma_2}(1-e^{-\gamma_2 t})$ is the solution of
the deterministic part of the equation, i.e. without the noise. On
using Eq. (\ref{gs2}), one can calculate the phenotypic noise at any
time, arbitrarily far from stationarity, and furthermore one can
study its behavior starting from an arbitrary initial amount of
proteins. Note that $\delta$ enters only through $\xi(t)$  and the
temporal evolution scales are determined by the transcription rate
and the degradation rate of proteins. Interestingly, one obtains a
distribution of proteins with a cut off along the interval
$[0,\xi_t)$ at any time whenever $\delta>0$. By exploiting the exact
dynamical solution one can study the evolution of the phenotypic
noise strength in finer detail and probe the experimental
consequences of the burst process hypothesis (Fig. (\ref{dyn})).

A measurable quantity that directly probes the protein distribution
and its temporal evolution is the cell-turnover distribution (CTD)
denoted by $\mathcal{P}(\rho,t)$ and defined as the probability that
at time $t$ the ratio $x(t)/x(0)$ is equal to $\rho$, where $x(t)$
and $x(0)$ are the number of proteins within an isogenic cell
population at time $t>0$ and $t=0$, respectively. This quantity can
be defined both close to and far from stationarity \cite{si}: if the
initial distribution of the proteins within the cell population is
in the steady state given by Eq. (\ref{ssng}) with $\delta=0$, then
at a subsequent time $t>0$ the CTD is

\begin{eqnarray}\label{ctd}
\mathcal{P}_{CTD}(\rho,t)&=& e^{-k_1 t}\delta\left(\rho-e^{-\gamma_2
t}\right)+\left(\frac{k_1}{\gamma_2}\right)^2\frac{e^{-k_1
t}(e^{\gamma_2 t}-1)\Theta\left(\rho-e^{-\gamma_2
t}\right)}{\left[1+e^{\gamma_2 t}(\rho-e^{-\gamma_2
t})\right]^{\frac{k_1}{\gamma_2}+1}}\times\nonumber\\
 & \times & {}_{2}F_1\left(\frac{k_1}{\gamma_2}+1,\frac{k_1}{\gamma_2}+1,2;
  \frac{(e^{\gamma_2 t}-1)(\rho-e^{-\gamma_2
t})}{1+e^{\gamma_2 t}(\rho-e^{-\gamma_2 t})}\right),
\end{eqnarray}

\begin{figure}[h]
\begin{center}
  \includegraphics[width=10cm]{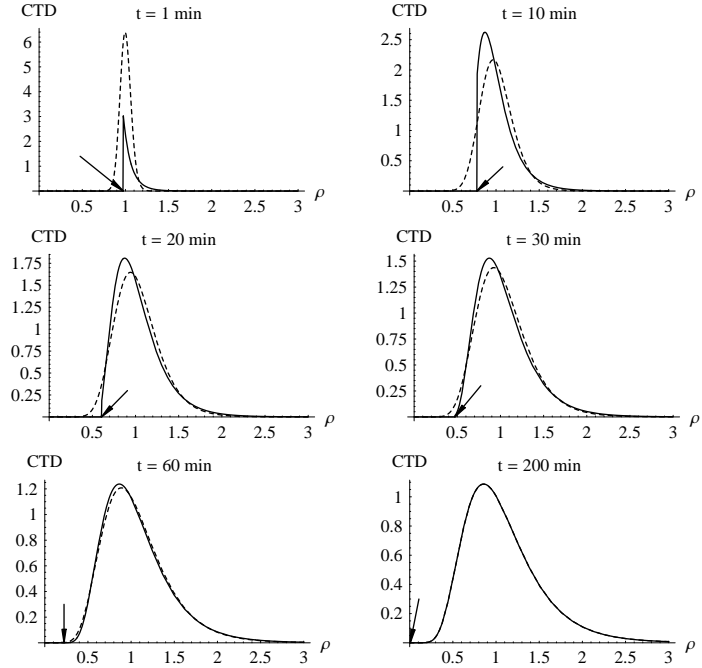}\\
  \caption{Cell Turnover Distribution (CTD). The dashed curve is for the
  Gaussian noise case,
  i.e. the CTD is calculated assuming that the governing equation is
  Eq. (\ref{gwn}) with
$f\equiv k_1 k_2/\gamma_1$ and $D\equiv \gamma_2 k_2/\gamma_1$;
  whereas the other curve is for the non-Gaussian noise case, i.e. for Eq. (\ref{ctd}).
   In this case
 the arrow indicates the cut-off point. Note, however, that
extrinsic noise could tend to smooth out the discontinuity. In both
cases $k_1/\gamma_2=12.88$ and we have set
  $\gamma_2^{-1}=40$ min.}\label{ctdfig}
\end{center}
\end{figure}

\no where  ${}_{2}F_1(a,b,c;x)$ is the standard hypergeometric
function \cite{Leb}. Thus, according to Eq. (\ref{ctd}), under the
burst process hypothesis we predict that \textit{i)} the CTD
vanishes between $0$ and $e^{-\gamma_2 t}$ even though the system is
at stationarity, an effect which ought to be detectable for time
scales less than or of the order of $1/\gamma_2$, \textit{ii)} the
CTD depends only on $k_1$ and $\gamma_2$ but not on translational
efficiency and other rates, \textit{iii)} at very large time
separation there is only one free parameter, the ratio
$k_1/\gamma_2$, and the CTDs predicted by the Gaussian and
non-Gaussian noises become the same (fig.(\ref{ctdfig})) \cite{si}.

The analogous time dependent solutions for the Gaussian white noise
can be compared with Eqs. (\ref{gs2}) and (\ref{ctd}) \cite{si}. The
closer the system is to its steady-state, the more difficult it is
to distinguish among the effects of gestation, senescence and
burst-like production. Thus an experimental protocol capable of
analyzing the cell population and its time evolution with different
initial conditions would be helpful to disentangle the nature of
stochastic noise. At early times, the evolution of the distribution
is strongly affected by the specific mechanisms involved in the
dynamics. At this stage, different distributions of waiting times
between events or burst sizes produce non-stationary distributions
that are very different, and the distinctive effects of noise,
deterministic driving forces or coupling of degrees of freedom can
be elucidated. Different conditions at initial times propagate into
the early temporal evolution in strongly different ways according to
the different effects of involved mechanisms, but inexorably lead to
the same distribution for large time separation.

We are grateful to Sunney Xie and his group for useful
correspondence.

\end{document}